\documentclass{INTERSPEECH2023}
\usepackage{multirow}
\usepackage[normalem]{ulem}
\useunder{\uline}{\ul}{}
\usepackage{booktabs}
\usepackage{tipa}
\usepackage{url}
\usepackage{subcaption}
\usepackage{diagbox}
\usepackage{cite}

\interspeechcameraready

\title{Incorporating Ultrasound Tongue Images for Audio-Visual Speech Enhancement through Knowledge Distillation}
\name{Rui-Chen Zheng, Yang Ai, Zhen-Hua Ling*\thanks{* Corresponding Author. This work was partially funded by the Fundamental Research Funds for the Central Universities.}}
\address{National Engineering Research Center of Speech and Language Information Processing, \\University of Science and Technology of China, Hefei, P. R. China}
\email{zhengruichen@mail.ustc.edu.cn,  \{yangai, zhling\}@ustc.edu.cn}

\begin{document}

\maketitle
 
\begin{abstract}
Audio-visual speech enhancement (AV-SE) aims to enhance degraded speech along with extra visual information such as lip videos, and has been shown to be more effective than audio-only speech enhancement. This paper proposes further incorporating ultrasound tongue images to improve lip-based AV-SE systems' performance. Knowledge distillation is employed at the training stage to address the challenge of acquiring ultrasound tongue images during inference, enabling an audio-lip speech enhancement student model to learn from a pre-trained audio-lip-tongue speech enhancement teacher model. Experimental results demonstrate significant improvements in the quality and intelligibility of the speech enhanced by the proposed method compared to the traditional audio-lip speech enhancement baselines. Further analysis using phone error rates (PER) of automatic speech recognition (ASR) shows that palatal and velar consonants benefit most from the introduction of ultrasound tongue images.
\end{abstract}
\noindent\textbf{Index Terms}: speech enhancement, audio-visual, ultrasound tongue image, knowledge distillation

\vspace{-1mm}
\section{Introduction}
\vspace{-1mm}
Speech enhancement (SE) is a crucial research problem in speech signal processing that aims to improve the quality and intelligibility of speech signals corrupted by various types of distortions \cite{kumar2016speech}. Numerous SE methods have been proposed to assist the application in several areas, such as speech or speaker recognition, hearing aids, and mobile communication \cite{chaudhari2015review}. Classical audio-only speech enhancement (AO-SE) approaches have been successful in estimating underlying target speech signals using either knowledge-based modeling \cite{ephraim1995signal,yang2005spectral}, or data-driven paradigms such as deep learning \cite{xu2014regression, defossez2020real}. Since speech perception is inherently multimodal, particularly audio-visual, recent studies have investigated using visual cues as well as acoustic signals for SE \cite{michelsanti2021overview}. This approach, known as audio-visual speech enhancement (AV-SE) \cite{almajai2010visually,hou2018audio, gabbay2018visual}, is more effective than simple AO-SE methods as visual cues are essentially unaffected by the acoustic environment.

Lip videos are the most commonly used visual cues for AV-SE due to their easy availability. Besides, they can also help disambiguate phonetically similar sounds since they record the movement of external articulators involved in speech production. A deep AV-SE model based on convolutional neural networks (CNNs) \cite{afouras2018conversation} was proposed to separate a speaker's voice by predicting both the magnitude and the phase of the target signal given lip regions. A novel framework \cite{xu2022vsegan} incorporated lip information for speech enhancement by integrating a generative adversarial network (GAN) to generate high-quality speech. Another study \cite{xu2022improving} proposed an AV-SE system that achieved impressive performance using a multi-layer fusion model with a multi-head cross-attention mechanism to fuse audio and lip features. 

Speech production is a complex process relying on multiple articulators, including the jaw, lips, teeth, and tongue. Using lip videos for speech processing tasks like SE often has limitations due to the lack of descriptions on internal articulators, e.g., tongue and velar. To address this issue, some studies have suggested employing internal articulation features captured using medical imaging techniques such as magnetic resonance imaging (MRI) \cite{scott2014speech} and ultrasound tongue imaging (UTI) \cite{stone2005guide}, to provide complementary data. Compared to MRI, UTI is relatively cheap, non-invasive, and can provide high-resolution images. UTI uses a real-time B-mode ultrasound transducer placed under the speaker’s chin to visualize a midsaggital or coronal view of the tongue during speech production. Ultrasound tongue images have been used in various speech processing tasks, such as speech recognition \cite{toth2018multi, ribeiro2021silent} and speech reconstruction \cite{porras2019dnn, zhang2021talnet}. However, their potential for speech enhancement tasks has yet to be fully explored, creating a research gap that could be addressed by leveraging the advantages of ultrasound tongue images. 

However, obtaining ultrasound tongue images is more complex than collecting lip videos due to the requirement of extra equipment. Therefore, this paper proposes introducing ultrasound tongue information to an audio-lip SE model through knowledge distillation. Specifically, an audio-lip-tongue SE teacher model with U-Net-based \cite{ronneberger2015u} structure is first proposed to incorporate ultrasound tongue images with lip videos to assist speech enhancement. An audio-lip SE student model is then guided to learn ultrasound tongue information from a pre-trained audio-lip-tongue SE teacher model via multiple loss functions. At inference time, the proposed audio-lip SE model is fed with noisy speech and lip videos, and the ultrasound tongue images are not necessary anymore. Experimental results demonstrate that the proposed method can generate speech with improved quality and intelligibility compared to conventional audio-lip baselines trained solely with clean speech supervision. Moreover, a notable reduction in phoneme error rate (PER), specifically for palatal and velar consonants, can be witnessed while applying automatic speech recognition (ASR) for transcription.

\begin{figure*}[t]
  \centering
  \setlength{\abovecaptionskip}{0.cm}
  \setlength{\belowcaptionskip}{-0.5cm}
  \includegraphics[width=0.95\linewidth]{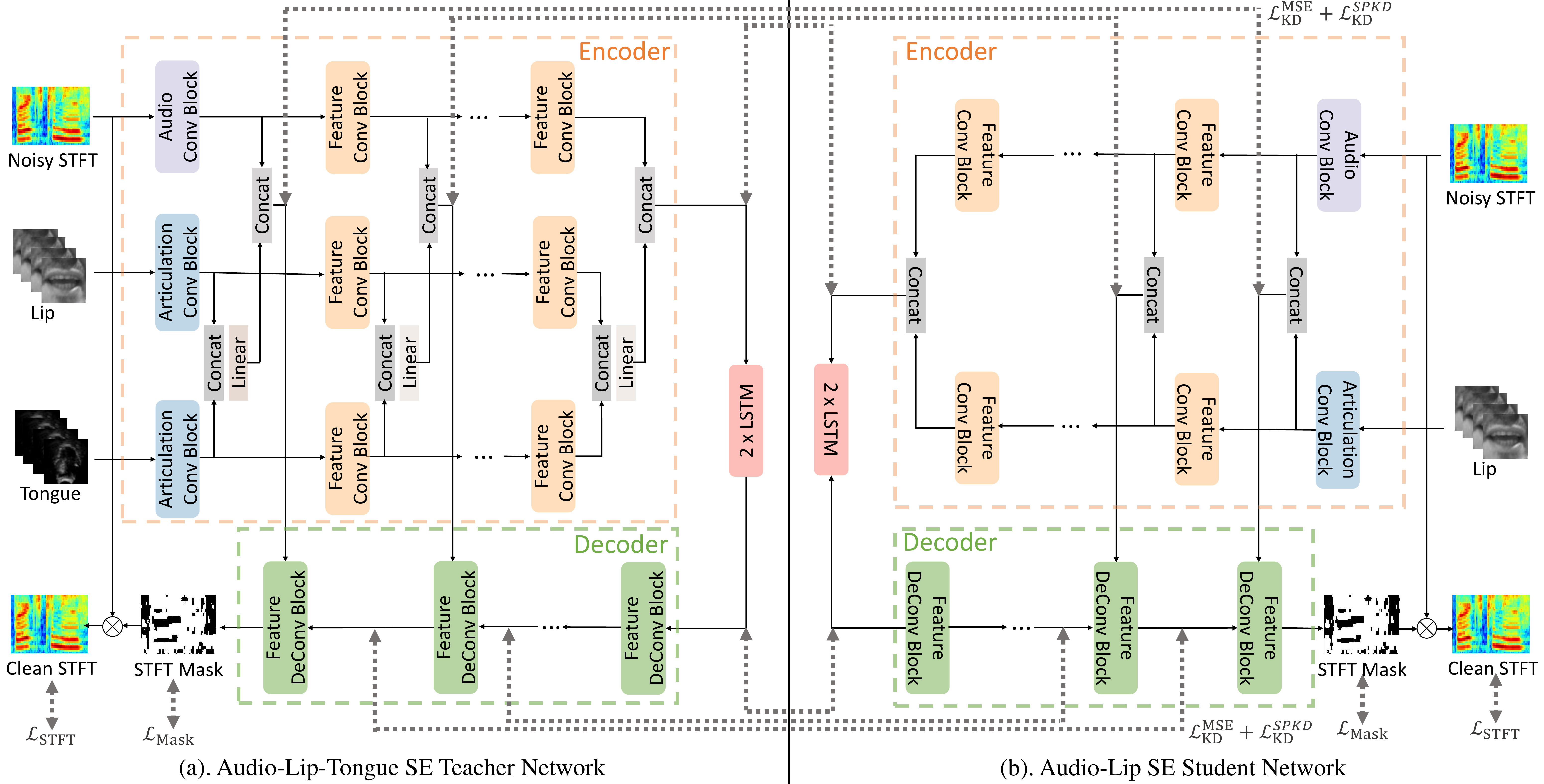}
  \caption{Details of the proposed method. The dashed arrows indicate the loss function. (a) shows the structure of the Audio-Lip-Tongue SE teacher model; (b) shows the structure of the Audio-Lip SE student model introducing ultrasound knowledge with knowledge distillation method. }
  \label{model}
  \vspace{-1mm}
\end{figure*}

\vspace{-1mm}
\section{Proposed Method}
\vspace{-1mm}
Under most circumstances, only a noisy speech and its corresponding lip video can be obtained as reference data since acquiring ultrasound tongue images is not as straightforward as capturing lip video. Hence, we propose enabling an audio-lip SE student model to assimilate ultrasound tongue information from a pre-trained audio-lip-tongue SE teacher model through knowledge distillation. Details are illustrated in Fig.~\ref{model} and will be introduced in this section. 

\vspace{-1mm}
\subsection{Audio-Lip-Tongue SE Teacher Model}
\vspace{-1mm}
Inspired by previous study \cite{gao2021visualvoice, xu2022vsegan}, we propose a U-Net \cite{ronneberger2015u} style teacher model with noisy speech, lip videos, and ultrasound tongue images as input as shown in Fig.\ref{model}a. The model can be roughly divided into a multi-modal encoder and a decoder connected by a long short-term memory (LSTM) based embedding block.

The encoder contains an articulation stream and an audio stream. For the articulation stream, we process both lip and tongue image sequences to obtain fused representations for describing articulation. Specifically, pixel-wise mean and standard deviation are computed for each utterance, repeated, and then appended as extra channels to the ultrasound and lip sequences \cite{zhang2021talnet}. The resulting input is of dimension $3\times T\times H\times W$, where $H$ and $W$ are the height and width of the lip and tongue images. The input is first processed by an articulation convolutional block consisting of three strided 3D-CNN layers, each followed by batch normalization and Leaky-ReLU. The resulting features are then flattened along the time axis and processed by seven feature convolutional blocks. Each block involves a series of 2D-CNN layers with frequency pooling layers to reduce the frequency while preserving the time dimension. The lip and tongue features are concatenated at each layer and then passed through a linear layer to obtain the fused articulation representations with reduced size.

For the audio stream, the encoder takes as input the real and imaginary parts of the noisy complex spectrogram with dimension $2\times F\times T$, where $F$ and $T$ are the frequency and time dimensions of the spectrogram. Each time-frequency bin contains the real and imaginary parts of the corresponding complex spectrogram value \cite{gao2021visualvoice}. The input is first processed by an audio convolutional block comprising two strided 2D-CNN layers and then by seven feature convolutional blocks as used in the articulation stream. 
Finally, the audio representation are concatenated with the fused articulation representations layer by layer to obtain the ultimate multi-modal representations for SE.

Two LSTM layers are inserted between the encoder and decoder to better model temporal dependencies. The decoder with skip connection exhibits a symmetric structure concerning the encoder, whereby the convolutional layer is substituted with an upconvolutional layer, and the frequency pooling layer is replaced by a frequency upsampling layer. The input of each decoder layer is the concatenation of the output of the previous layer and the multi-modal representations given by the corresponding encoder layer. The final output feature map is fed through an activation layer to predict a complex mask with the same dimensions as the input noisy spectrogram. The resulting mask is applied to the noisy input through complex multiplication, yielding an enhanced complex spectrogram which is transformed back into the time domain via an inverse short-time fourier transform (iSTFT). 

The training criterion of the teacher model is to minimize the following loss
\begin{equation}
    \setlength{\abovedisplayskip}{3pt}
    \setlength{\belowdisplayskip}{3pt}
    \mathcal{L}_{Teacher} = \mathcal{L}_{Mask} + \alpha \mathcal{L}_{STFT},
    \label{Lse}
\end{equation}
where $L_{Mask}$ and $L_{STFT}$ denote the mean square error (MSE) losses of the mask and the complex spectrogram, respectively. The hyper-parameter $\alpha$ is utilized to ensure both $L_{Mask}$ and $L_{STFT}$ are scaled to the similar magnitude.

\vspace{-1mm}
\begin{table*}[htbp]
\setlength{\abovecaptionskip}{0.cm}
\setlength{\belowcaptionskip}{-0.5cm}
\renewcommand\arraystretch{0.95}
\caption{Evaluation results of the proposed model compared with baseline models. Best results are highlighted in \textbf{bold}.}
\label{Table 1}
\centering
\begin{tabular}{cc|lll|lll|lll}
\toprule
\toprule
\multicolumn{1}{c|}{\multirow{2}{*}{Method}}      & SNR          & \multicolumn{3}{c|}{2.5dB}                                                        & \multicolumn{3}{c|}{-2.5dB}                                                       & \multicolumn{3}{c}{-7.5dB}                                                       \\ \cline{2-11} 
\multicolumn{1}{c|}{}                            & Metrics      & \multicolumn{1}{c}{SegSNR} & \multicolumn{1}{c}{PESQ} & \multicolumn{1}{c|}{STOI} & \multicolumn{1}{c}{SegSNR} & \multicolumn{1}{c}{PESQ} & \multicolumn{1}{c|}{STOI} & \multicolumn{1}{c}{SegSNR} & \multicolumn{1}{c}{PESQ} & \multicolumn{1}{c}{STOI} \\ \hline
\multicolumn{2}{c|}{Noisy}   & 0.2414                     & 1.4285                    & 0.8503                    & -3.1703                     & 1.2283                   & 0.7527                    & -5.9847                     & 1.1298                    & 0.6229                   \\ \hline
\multicolumn{2}{c|}{Gabby et al. 2018 \cite{gabbay2018visual}}   & 4.8731                     & 1.8260                    & 0.8871                    & 3.8721                     & 1.6502                   & 0.8566                    & 2.6978                     & 1.4590                    & 0.8005                   \\
\multicolumn{2}{c|}{Hegde et al. 2021 \cite{hegde2021visual}}    & 3.6086                     & 1.8756                   & 0.8882                    & 2.9790                      & 1.6890                    & 0.8603                    & 2.0987                     & 1.4863                   & 0.8052                   \\
\multicolumn{2}{c|}{Hussain et al. 2021 \cite{hussain2021towards}} & 6.1654                     & 2.0388                   & 0.9245                    & 4.5055                     & 1.7750                    & 0.8867                    & 2.8217                     & 1.5424                   & 0.8177                   \\
\multicolumn{2}{c|}{Proposed}                                   & \textbf{9.0061}            & \textbf{2.1073}          & \textbf{0.9335}           & \textbf{6.8343}            & \textbf{1.8241}          & \textbf{0.8911}           & \textbf{4.6944}            & \textbf{1.5858}          & \textbf{0.8221}         \\
\bottomrule
\bottomrule
\end{tabular}
\vspace{-1mm}
\end{table*}

\vspace{-1mm}
\subsection{Audio-Lip SE Student Model}
\vspace{-1mm}

An audio-lip SE student model is proposed with the architecture depicted in Fig.\ref{model}b, where the input of tongue images is removed compared with the teacher. To utilize tongue information even in the absence of ultrasound tongue images during reference, we employ knowledge distillation to train the student. Specifically, in addition to the supervised training with the backbone losses $L_{Mask}$ and $L_{STFT}$ described in Eq.(\ref{Lse}), the student model also receives supervision signals from the teacher. During forward inference of the student and teacher networks, the output of each layer is saved to compute the knowledge distillation loss for each layer of the encoder, decoder, and LSTM embedding blocks. For each layer, the MSE loss between the output features of the teacher and the student is calculated following
\begin{equation}
 \setlength{\abovedisplayskip}{3pt}
 \setlength{\belowdisplayskip}{3pt}
 \mathcal{L}_{KD}^{MSE} = \sum_{l=1}^{N} ||\bm{F}_T^l- \bm{F}_S^l||_2^2,
 \label{L_KD_MSE}
\end{equation}
where $N$ represents the number of layers, and $\bm{F}_T^l$ and $\bm{F}_S^l$ denote the output features of the $l^{th}$ layer of the teacher and student model, respectively. We guarantee that the feature dimensions of the corresponding layers in the teacher and student models are identical.

The similarity-preserving knowledge distillation (SPKD) loss \cite{tung2019similarity, cheng2022cross} is utilized to further supervise the student model. The SPKD loss aims to achieve dimensional compression and to transmit similarity information simultaneously by computing pairwise similarity matrices. Given a mini-batch input, the feature map of the $l^{th}$ layer is represented as $\bm{F}^l \in \mathbb{R}^{b\times c\times t\times f}$, where $b$ is the batch size, $c$ is the number of output channels, $t$ is the number of frames, and $f$ is the dimension of the feature space. To account for potential interference of information from different frames, the features are first segmentated on frame-level and then flattened into two dimensions. The transformed feature of the $j^{th}$ frame is denoted as $\bm{F}^{(l,j)} \in \mathbb{R}^{b\times f'}$, where $f'=c\times f$. The similarity matrice of each frame for the teacher $\bm{G}_T^{(l,j)}$ and the student $\bm{G}_S^{(l,j)}$  is calculated independently following \cite{cheng2022cross}. The SPKD loss for the $l^{th}$ layer is then calculated as the sum of the similarity distances across all frames. The overall SPKD loss is determined by summing the SPKD loss of each layer as follows
\begin{equation}
\setlength{\abovedisplayskip}{3pt}
\setlength{\belowdisplayskip}{3pt}
\mathcal{L}_{KD}^{SPKD} = \sum_{l=1}^{N}\mathcal{L}_{SPKD}^l = \frac{1}{b^2}\sum_{l=1}^{N}\sum_{j=1}^T ||\bm{G}_T^{(l,j)}- \bm{G}_S^{(l,j)}||_F^2,
\label{L_KD_SKD}
\end{equation}
where $||\cdot||_F$ is the Frobenius norm.

Therefore, to train the proposed audio-lip SE model, the overall loss function can be written as
\begin{equation}
\setlength{\abovedisplayskip}{3pt}
\setlength{\belowdisplayskip}{3pt}
\mathcal{L}_{Student} = \mathcal{L}_{Mask} + \alpha \mathcal{L}_{STFT} + \gamma_1\mathcal{L}_{KD}^{MSE} + \gamma_2\mathcal{L}_{KD}^{SPKD},
\label{L_student}
\end{equation}
where, $\alpha, \gamma_1$ and $\gamma_2$ are all hyper-parameters used to ensure that the loss functions are scaled to the similar magnitude.

\vspace{-1mm}
\section{Datasets and Implementation Details}

\subsection{Datasets}
\vspace{-1mm}
For our experiments, we utilized the TaL corpus \cite{ribeiro2021tal}, a multi-speaker dataset containing ultrasound tongue imaging, optical lip video, and audio for each utterance. Specifically, we employed 10,271 utterances from 73 speakers for training, and 810 utterances from 81 speakers for validation, where eight speakers from the validation sets were unseen in the training set. The test set included 1749 utterances from 73 seen speakers and 1407 utterances from 8 unseen speakers. The content of the three sets was mutually exclusive from each other.

Noisy speech was generated by mixing noise with the clean speech in the TaL corpus. During training, we introduced ten different noise types as in \cite{valentini2018speech}: eight noise recordings from the DEMAND database \cite{thiemann2013diverse} and two artificially generated noises, namely speech-shaped noise and babble noise. These noise types were added to the speech signal at three different signal-to-noise (SNR)  values (0dB, -5dB, and -10dB). For validation and testing, we added five additional noises: living room, office, bus, street cafe, and a public square. We used slightly higher SNR values (2.5dB, -2.5dB and -7.5dB) than the ones used during training following previous work \cite{valentini2018speech}.

\vspace{-1mm}
\subsection{Implementation Details}
\vspace{-1mm}
The lip videos and ultrasound tongue images were processed following the pipeline described in \cite{zhang2021talnet}. The resulting lip videos and ultrasound tongue images were resized to $64\times128$ and had a frame rate of 81.5 fps. For audio preprocessing, the audio was downsampled to 16 kHz, and the STFT was computed using a Hann window with a length of 512, a hop size of 196 and an FFT point number of 512 to match the frame rate of the ultrasound. The resulting complex spectrogram had a dimension of $2\times257\times T$.

We employed the Adam optimizer with an initial learning rate of 1e-3 for backpropagation. The learning rate decreased by 0.1 once learning stagnated, i.e., the validation error did not improve for ten epochs. The model was trained at utterance level by randomly cropping all the samples in the mini-batch to have the same number of frames as the shortest one.

\vspace{-1mm}
\begin{table*}[]
\setlength{\abovecaptionskip}{0.cm}
\setlength{\belowcaptionskip}{-0.5cm}
\renewcommand\arraystretch{0.95}
\caption{Evaluation results of the proposed model in ablation studies. Best results are highlighted in \textbf{bold}. \uline{Underline} characters indicate the sub-optimal results.}
\label{Table 2}
\centering
\begin{tabular}{cc|ccc|ccc|ccc}
\toprule
\toprule
\multicolumn{1}{c|}{\multirow{2}{*}{Method}} & SNR     & \multicolumn{3}{c|}{2.5dB}                                                        & \multicolumn{3}{c|}{-2.5dB}                                                       & \multicolumn{3}{c}{-7.5dB}                                                       \\ \cline{2-11} 
\multicolumn{1}{c|}{}                       & Metrics & \multicolumn{1}{c}{SegSNR} & \multicolumn{1}{c}{PESQ} & \multicolumn{1}{c|}{STOI} & \multicolumn{1}{c}{SegSNR} & \multicolumn{1}{c}{PESQ} & \multicolumn{1}{c|}{STOI} & \multicolumn{1}{c}{SegSNR} & \multicolumn{1}{c}{PESQ} & \multicolumn{1}{c}{STOI} \\ \hline
\multicolumn{2}{c|}{Proposed}                         & {\ul 9.0061}               & 2.1073                   & {\ul 0.9335}              & {\ul 6.8343}               & {\ul 1.8241}             & {\ul 0.8911}              & {\ul 4.6944}               & {\ul 1.5858}             & {\ul 0.8221}             \\
\multicolumn{2}{c|}{w/o SPKD}                         & 8.9927                     & \textbf{2.1133}          & 0.9328                    & 6.8251                     & 1.8218                   & 0.8892                    & 4.6442                     & 1.5802                   & 0.8194                   \\
\multicolumn{2}{c|}{w/o KD (Audio-Lip)}                           & 8.9741                     & 2.0841                   & 0.9304                    & 6.8073                     & 1.8030                    & 0.8854                    & 4.6314                     & 1.5604                   & 0.8142                   \\ \hline
\multicolumn{2}{c|}{Audio-Lip-Tongue}                 & \textbf{9.0315}            & {\ul 2.1122}             & \textbf{0.937}            & \textbf{6.9267}            & \textbf{1.8322}          & \textbf{0.8987}           & \textbf{4.8707}            & \textbf{1.5973}          & \textbf{0.8394}          \\
\multicolumn{2}{c|}{Audio-Tongue}                     & 8.8509                     & 2.0990                    & 0.9321                    & 6.7223                     & 1.8133                   & 0.8901                    & 4.6153                     & 1.5631                   & 0.822                    \\
\multicolumn{2}{c|}{Audio-Only}                       & 8.6748                     & 1.9872                   & 0.9215                    & 6.4224                     & 1.7088                   & 0.8661                    & 4.0888                     & 1.4773                   & 0.7739      \\
\bottomrule
\bottomrule
\end{tabular}
\end{table*}

\vspace{-1mm}
\begin{table*}[]
\renewcommand\arraystretch{0.94}
\setlength{\abovecaptionskip}{0cm}
\setlength{\belowcaptionskip}{-0.3cm}
\caption{PERs (\%) of different phoneme categories for clean speech, noisy speech, and the speech enhanced by different methods. Consonant phonemes are divided according to the place of articulation. Best results are highlighted in \textbf{bold}. \uline{Underline} characters indicate the sub-optimal results. The numbers in parentheses show relative PER reduction (\%) compared with the audio-only method.}
\label{PER}
\begin{tabular}{c|c|c|cccccccc}
\toprule
\toprule
\multirow{2}{*}{Methods} & \multirow{2}{*}{Silence} & \multirow{2}{*}{Vowels} & \multicolumn{8}{c}{Consonants}                                                                                                                                                                                           \\ \cline{4-11} 
                         &                          &                         & Labial         & \begin{tabular}[c]{@{}c@{}}Labio-\\ dental\end{tabular} & Dental         & Alveolar       & \begin{tabular}[c]{@{}c@{}}Alveo-\\ palatal\end{tabular} & Palatal        & Velar          & Glottal        \\ \hline
Clean                    & 0.58                     & 3.09                    & 2.54           & 1.24                                                    & 2.18           & 2.96           & 1.94                                                     & 1.88           & 2.69           & 2.46           \\ 
Noisy                    & 25.10                    & 46.10                   & 41.58          & 37.44                                                   & 43.43          & 41.74          & 33.03                                                    & 37.53          & 37.10          & 46.35          \\ \hline
Audio-Only               & 10.54                    & 31.97                   & 33.54          & 34.03                                                   & 31.02          & 33.46          & 14.8                                                    & 30.78          & 27.37          & 33.76          \\  \hline
Audio-Lip                & 7.71                     & 27.67                   & 27.97          & {\ul 31.82}                                                   & {\ul 24.78}    & 30.29          & 12.48                                               & 30.33          & 26.48          & 30.42          \\
 & (26.85)                   & (13.45)                  & (16.61)         & (6.49)                                                  & (20.12)         & (9.47)          & (15.68)                                                   & (1.46)          & (3.25)          & (9.89)          \\   \hline
Audio-Lip-Tongue         & \textbf{6.34}            & \textbf{20.22}          & \textbf{24.95} & 32.15                                       & \textbf{22.40} & \textbf{25.61} & \textbf{11.13}                                            & \textbf{23.55} & \textbf{17.35} & \textbf{25.09} \\
 & (39.84)                   & (36.75)                  & (25.61)         & (5.52)                                                   & (27.79)         & (23.46)         & (24.80)                                                   & (23.49)         & (36.61)         & (25.68)         \\  \hline
Proposed                 & {\ul 7.34}               & {\ul 25.31}             & {\ul 26.42}    & \textbf{30.56}                                             & 25.44          & {\ul 29.35}    & {\ul 11.17}                                                     & {\ul 26.59}    & {\ul 22.85}    & {\ul 29.87}    \\
 & (30.36)                   & (20.83)                  & (21.23)         & (10.20)                                                  & (17.99)         & (12.28)         & (24.53)                                                   & (13.61)         & (16.51)         & (11.52)        \\
\bottomrule
\bottomrule  
\end{tabular}
\end{table*}

\section{Experimental Results}

\subsection{Overall Performance}
\vspace{-1mm}

Three recent AV-SE methods \cite{gabbay2018visual, hegde2021visual, hussain2021towards} whose source codes were available online were used as baselines for comparison with our proposed audio-lip SE model. When evaluating the method \cite{hegde2021visual}, as a generated pseudo lip stream was adopted for AV-SE in its original paper, we used the natural lip stream for a fair comparison. Each baseline was trained on our constructed dataset following the training strategy in its original paper. Segmental signal-to-noise ratio (SegSNR), perceptual evaluation of speech quality (PESQ), and short-term objective intelligibility (STOI) were used as evaluation metrics. The results in Table 1 demonstrate that our proposed method outperforms all three baselines in all evaluation metrics, indicating improved speech quality and intelligibility\footnote{Speech samples are available at: \url{https://zhengrachel.github.io/UTIforAVSE-demo/}}. 

\vspace{-1mm}
\subsection{Ablation Studies}
\vspace{-1mm}

A series of ablation studies were conducted to provide further evidence of the efficacy of incorporating ultrasound tongue knowledge through the knowledge distillation method. The outcomes are presented in Table 2. Specifically, the effectiveness of the SPKD loss was demonstrated by removing the loss function $\mathcal{L}_{KD}^{SPKD}$ in Eq.\ref{L_student} (\textit{``w/o SPKD"}). To demonstrate the effectiveness of introducing ultrasound tongue information via the knowledge distillation method, an audio-lip SE model was trained solely with $\mathcal{L}_{Mask}$ and $\mathcal{L}_{STFT}$ (\textit{``w/o KD (Audio-Lip)"}). Additionally, for reference, we included results from the well-trained audio-lip-tongue SE teacher model together with the audio-tongue and audio-only SE models built by removing the lip or the whole articulation inputs in the teacher model. 

Table 2 shows that while the audio-lip-tongue SE teacher model achieved the best performance, the proposed method won the second place on most metrics. \textit{``w/o SPKD"} achieved results inferior to the proposed method except for SNR=2.5dB, indicating the effectiveness of introducing the SPKD loss particularly in low SNR scenarios. The outcomes of the proposed method surpassed those of the \textit{``w/o KD (Audio-Lip)"}, suggesting that the knowledge distillation method effectively incorporates a portion of the articulation knowledge derived from the ultrasound tongue. Furthermore, the performance of the proposed method outperformed that of both the audio-tongue and audio-only models, highlighting the effectiveness of combining multi-modal sources of articulation knowledge.

\vspace{-1mm}
\subsection{Analysis on Phoneme Categories}
\vspace{-1mm}
To further study the gains of incorporating ultrasound tongue images, we employed an ASR engine to analyze the enhanced speech's PERs of different phoneme categories. An ASR API provided in ESPNet\footnote{\url{https://github.com/espnet/espnet_model_zoo}} \cite{watanabe2018espnet} was utilized to transcribe the enhanced speech. Each enhanced utterance was forcibly aligned with its transcription by the Montreal Forced Aligner (MFA)\footnote{\url{https://github.com/MontrealCorpusTools/Montreal-Forced-Aligner}} \cite{mcauliffe2017montreal} tool, consequently obtaining a time-aligned phoneme sequence. Ground truth phoneme sequences were obtained by aligning clean utterances with their corresponding texts using the MFA tool. The recognized and ground truth phoneme sequences were further aligned by minimizing the edit distance, and the PERs for different phoneme categories could be calculated. Phonemes were categorized according to the English (UK) MFA dictionary v2\_0\_0\footnote{\url{https://mfa-models.readthedocs.io/en/latest/dictionary/English/}}. 

The results are shown in Table 3. Comparing audio-lip and audio-only methods, we can see that after introducing lip information for SE, PERs reduced significantly for some lip-related phonemes, such as silence, labials and dentals. However, for palatal and velar consonants, the relative PER reductions were quite small which indicates the limitations of using  only lip videos as articulation information. After further incorporating ultrasound tongue images, the audio-lip-tongue method achieved much more uniform PER reductions among phoneme categories than the audio-lip method. For palatals, velars and vowels whose articulations were mainly determined by internal tongue movement, their relative PER reductions improved significantly from 1.46\% to 23.49\%, from 3.25\% to 36.61\% and from 13.45\% to 36.75\%, respectively. Although not using tongue images at inference time, our proposed method also achieved lower PERs than the audio-lip method for most phoneme categories, especially for palatals, velars and vowels, indicating the effectiveness of incorporating tongue information through knowledge distillation.

\vspace{-1mm}
\section{Conclusion}
\vspace{-1mm}
This paper proposes incorporating ultrasound tongue images via knowledge distillation for AV-SE systems. An audio-lip-tongue SE model is first trained and used as a teacher to guide the training of an audio-lip SE student model through knowledge distillation. Our proposed method demonstrates superior performance over the conventional AV-SE baselines. Furthermore, the evaluated PER results indicate that lip and tongue modalities provide valuable articulation knowledge. Investigating the feasibility of constructing pseudo ultrasound tongue features from audio and lip video modalities to assist speech processing tasks will be our future work. 

\bibliographystyle{IEEEtran}
\bibliography{mybib}

\end{document}